\documentstyle{camera}
\newcommand{\AmS}{{\protect\the\textfont2
  A\kern-.1667em\lower.5ex\hbox{M}\kern-.125emS}}

\newcommand{\beq}{\begin{equation}}
\newcommand{\eeq}{\end{equation}}
\newcommand{\bea}{\begin{eqnarray}}
\newcommand{\eea}{\end{eqnarray}}

\def\dm2{\Delta m^2}
\def\sq2{sin^2(2\Theta)}

\input epsf

\begin{document}

%%%%%%%%%%%%%%%%%%%%%%%%%%%%%%%%%%%%%%%%%%%%%%%%%%%%%%%%
% The title, all uppercase; if you want to split it in
% two or more lines, put a \\ macro at each line break
% example:
%   \title{TITLE: FIRST LINE\\ SECOND LINE}
%
\title{A unique mechanism  generating the knee and the ankle
in the local galactic zone}

%%%%%%%%%%%%%%%%%%%%%%%%%%%%%%%%%%%%%%%%%%%%%%%%%%%%%%%%
% The Author(S), Separated By Commas; Do not put a
% comma before the last author, use instead the \And
% macro which produces a normal ``and'' in the
% caps/small caps context
%
\author{ANTONIO CODINO$^1$ AND FRANCOIS PLOUIN$^2$}

%%%%%%%%%%%%%%%%%%%%%%%%%%%%%%%%%%%%%%%%%%%%%%%%%%%%%%%%
%
\organization{$^{1)}$ Dipartimento di Fisica dell'Universita
         di Perugia and INFN, Italy.\\
$^{2)}$ former CNRS researcher, LLR,
         Ecole Polytechnique, F-91128 Palaiseau,France.}

\maketitle

\begin{abstract}
A new, unique mechanism accounting for the knee and the ankle in the
energy spectrum of the cosmic radiation is presented. 
The interplay of
the form and strength of the galactic  magnetic field, 
the rising trend of the nuclear cross sections
with energy, the position of the solar cavity within the Galaxy and disc size 
generates
 knees and ankles of individual
ions. The influence of these
observational data on the cosmic-ray intensity at Earth
is determined  by the appropriate simulation of the  cosmic-ray trajectories 
in the galactic volume.
The solid and extensive accord between the
computed and measured energy spectra of individual ions and of all ions
is  discussed and emphasized.
\end{abstract}
%%%%%\vspace{1.0cm}

\section{Introduction}
The differential energy spectrum of cosmic rays  presents
 two small but significant deviations, called knee and ankle, at the nominal energies
of $3 \times 10^{15}$ $eV$ and $6 \times 10^{18}$ $eV$,  respectively. 
After the first observation of the knee in 1958 (Kulikov et al., 1958)
 many experiments confirmed, over the decades,  the structure of the knee characterized 
by a specific bend energy, at about  $3 \times 10^{13}$ $eV$,  and a 
spectral index of 3,  above $6 \times 10^{15}$ $eV$.
Only in recent years, experimentally, has an important step 
forward been taken by the Kascade Collaboration (Kampert et al., 2003; 
Haungs et al.,2004; Antoni et al., 2005)
which identified and measured  
the knees of  individual ions in a very large energy band ($10^{15}$-$10^{17}$ $eV$),  
with relatively small error bars, and accurately charting the characteristic
 forms of the energy spectra. A vast literature (see,
 for example, Erlykin, 1996)
 testifies of the many
 attempts to explain and clarify these two deviations, over almost fifty years.

One notable aspect of the solution of the knee problem described here is that the
 explanation of the ankles
 of individual ions comes as a gratuity, once the knees are accounted for.

\par Prerequisites to comprehend the 
present study may be found  in a  series of previous works regarding the
 method of calculation (Codino, Brunetti and Menichelli,  1995;  
  Brunetti and Codino, 2000, Paper I) and 
the notion  of galactic basin (Codino and Plouin, 2003; 
Codino and Plouin, 2006, Paper II)
largely employed in the present study. 
Section 7 describes  an important, new inference, based on observational
facts,   which relates the 
knee energy to the ankle energy of the individual ions.
This inference is a cross-check, independent from the method of calculation
and the notion of galactic basin,  that the
mechanism generating the  ankle and the knee described here is correct. 
\par Due to the limited size of the
present paper some collateral
 aspects 
of the explanation of the knees are omitted; among
these: 
(1) how  the
agreement with the experimental data is affected by the particular values
 of the spectral indices and elemental composition
of cosmic ions.
(2) Implications and justification of the hypothesis of adopting constant
    spectral indices over the entire energy interval of the cosmic 
radiation, i.e.  $10^{10}$-$10^{21}$ $eV$.
(3) Direct, important consequences of this solution of the knee problem
      in our discipline. For instance, according to this study, any acceleration
mechanisms in the Galaxy are extraneous to the origin of the knees and  ankles.
Another notable consequence is that the extragalactic component of cosmic rays 
is likely to be marginal even 
at $10^{20}$ $eV$.
(4) Bibliography of previous attempts to explain the knee reported elsewhere
    (Codino and Plouin, 2006 Paper III).

\section{Evaluating cosmic ray intensity using cosmic ray trajectories}

In order to evaluate the cosmic ray intensity in a given small volume of the
Galaxy, for example the solar cavity, the number of cosmic ray trajectories 
intercepting it has to be computed. Trajectories are just the physical paths 
of cosmic rays determined by computer simulation. The initial point of any
trajectory is the source while the final point is the site where the 
cosmic ion disappears from the disk. Nuclear collisions, ionization energy
losses and escape from the disc are the three causes that make cosmic rays
disappear from the disc.
 %%%%%%%%%%%%%%%%%%%%%%%%%%%%%%%%%%%%%%%%%%%%%%%%%%%%%%%%%%%%%%%%
%%%%%\begin{figure}[t!]  %%% FIGURE 1 %%%
\begin{figure} [b!] %%% FIGURE 1 %%%
\vspace{-1.0cm}
\epsfysize=8cm \hspace{1.8cm}
%%%%%\epsfbox{fig02-gin-champ-spiral.eps}
\epsfbox{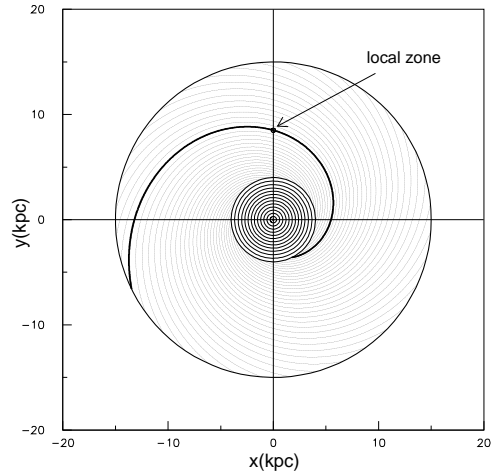}
%%%%%\vspace{0.3cm}
\vspace{-1.0cm}
\caption[h]{ The regular component of the magnetic
field in the disk is spiral according to a variety of measurements
in our own Galaxy and many other spiral galaxies morphologically similar
to the Milky Way.
The solid black curve denotes the principal magnetic field line; it is rooted on
the bulge edge at $r$=$4$ $kpc$ terminating at $r$=$15$ $kpc$ with
a length of $40.7$ $kpc$.  This
particular spiral intercepts the solar cavity at a distance of 8.5 $kpc$
from the galactic center. }
\end{figure}
%%%%%%%%%%%%%%%%%%%%%%%%%%%%%%%%%%%%%%%%%%%%%%%%%%%%%%%%%%%%%%%%

%%%%%% Table 1 %%%%%%%%%%%%%%%%%%%%%%%%%%%%%%%%%%%%%%%%%%%%%%%%%%%%%%%%%%%%
\begin{table}[t!]
\vspace{-0.5cm}
\begin{center}
\caption{Shapes and field strengths of the regular magnetic fields in spiral galaxies
determined by radio continuum measurements.} 
%%%%%\bigskip
\begin{tabular}{llll}
\hline
Galaxy      & Regular field      & field strength & References            \\
            & Shape              & $\mu G$        &                       \\
\hline
M33         & spiral             & $ 3 \pm 1 $    & Sofue et al. 1981     \\ 
M51         & spiral             &  10            & Mathewson et al. 1972 \\
NGC6946     & spiral             & $12 \pm 4$     & Klein et al. 1982     \\
M81         & spiral             & $ 8 \pm 3$     & Sofue et al. 1980     \\
M83         & spiral             &  not given     & Ondrechen 1985        \\
M31         & circular           &   3-4          & Beck 1981             \\
IC342       & circular           & $ 7 \pm 2$     & Graeve et al. 1988    \\
NGC253      & probab. spir.      &  a few $\mu G$ & Klein et al. 1083     \\
NGC2903     & spiral             &  not given     & Sofue et al. 1985     \\
NGC5055     & probab. spir.      &  not given     & Sofue et al. 1985     \\
NGC4631     & not given          &  a few $\mu G$ & Golla et al. 1988     \\
NGC891      & parall. gal. plane &  a few $\mu G$ & Sukumar et al. 1991   \\
NGC4565     & parall. gal. plane &  not given     & Sukumar et al. 1991   \\
NGC3251     & spiral             & $10$ $\mu G$ & Knapik et al. 2000    \\
NGC5055     & spiral             & $9$ $\mu G$ & Knapik et al. 2000    \\
\hline
\end{tabular}
\end{center}
\vspace{-0.5cm}
\end{table}
%%%%%%%%%%%%%%%%%%%%%%%%%%%%%%%%%%%%%%%%%%%%%%%%%%%%%%%%%%%%%%%%%%

\par The following astronomical, astrophysical and radioastronomy observations, 
are incorporated in the 
computational algorithms:
(1) the spiral field pattern shown in figure 1;
(2) the magnetic field strength (see figure 4 in  Paper II);
(3) the form and the dimension of the Galaxy (see
    figure 1 in Paper I); 
(4) a uniform distribution of cosmic ray sources in the
    galactic disk (see eqn. (5) in Paper I);  
(5) the nuclear cross sections ion-hydrogen, $\sigma$;
(6) the interstellar matter density  in the disk, $d$,  of 1.24 hydrogen 
atoms per $cm^3$ (Gaisser, 1990);   
 (7) the position of the solar cavity inside the 
disk, shown in figure 1, at 14 $pc$ from the galactic midplane. 
\par A cosmic ion emanated from a source travels, on average,  a mean length $L$, 
in the disk volume, primarily determined by $\sigma$,
the magnetic field structure and $d$. Typically, this global
length $L$ is subdivided into thousands and thousands of segments, depending on
the particular region of propagation and the energy of the cosmic ion. The ion 
propagation normal to the regular spiral field is generated by the chaotic
field leading to a transverse-to-longitudinal displacement ratio of 0.031,
value compatible with the quasi-linear theory of ion propagation in an
astrophysical environment (Giacalone $\&$ Jokipii, 1999).
The chaotic magnetic field
is materialized by magnetic cloudlets
(see, for instance, Vall\'ee, 1998)  with variable dimensions
and a field strentgh a factor 3
greater than that of the regular field.

\section{Galactic basins around the knee energy region}

 Most of the cosmic ions wandering in the Galaxy 
do not intercept any given recording instrument, placed in a specified site in 
the Galaxy,  just because of an abnormal distance source-instrument, 
or possibly, because of the low probability of encounter, due to the gas 
density, high nuclear cross sections or magnetic barriers. 
Generally, only a subset of the cosmic rays emanated by the
galactic sources intercepts any given recording
instrument.
In figure 2 this notable aspect is illustrated.
%%%%%%%%%%%%%%%%%%%%%%%%%%%%%%%%%%%%%%%%%%%%%%%%%%%%%%%%%%%%%%%%
\begin{figure}[t!]  %%% FIGURE 2 %%%
\epsfysize=9cm
\hspace{1.4cm}
%%%%%\epsfbox{fig11-b-gin-fer-haint-clp.eps}
\epsfbox{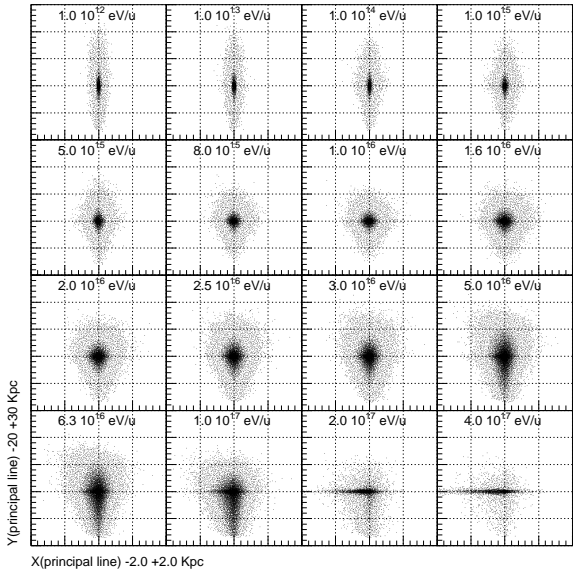}
%%%%%\vspace{0.3cm}
\vspace{-0.8cm}
\caption[h] { Source distributions for iron nuclei projected onto the 
galactic midplane
 at sixteen different energies,  from one $TeV$/$u$
up to the ankle energy region. The spiral reference frame, bound to the 
the principal magnetic field line, is used. Note the different
scales in the two axes. The disruption of the source pattern,  
going from an oval to a squat form as the energy increases, is obvious.}
\end{figure}
%%%%%%%%%%%%%%%%%%%%%%%%%%%%%%%%%%%%%%%%%%%%%%%%%%%%%%%%%%%%%%%%   
The positions of the sources emitting cosmic rays reaching the
local zone are far from being uniformly distributed in space; on the contrary, 
they accumulate in particular disk regions, generally in the vicinity of the
instrument. In figure 2, sixteen source distributions of iron nuclei, projected onto the 
galactic  midplane, are given for sixteen different energies,
from $5.6 \times 10^{14}$ $eV$ up to the ankle energy region.
 
%%%%%%%%%%%%%%%%%%%%%%%%%%%%%%%%%%%%%%%%%%%%%%%%%%%%%%%%%%%%%%%%
\begin{figure}[t!]  %%% FIGURE 3 %%%
%%%%%\epsfysize=10cm \hspace{1.0cm}
%%%%%\epsfbox{fig07-gin-fe-ariete.eps} \vspace{0.3cm}
\epsfysize=6cm \hspace{0.5cm}
%%%%%\epsfbox{fig07-gin-fe-ariete-z21.eps}
\vspace{0.3cm}
\epsfbox{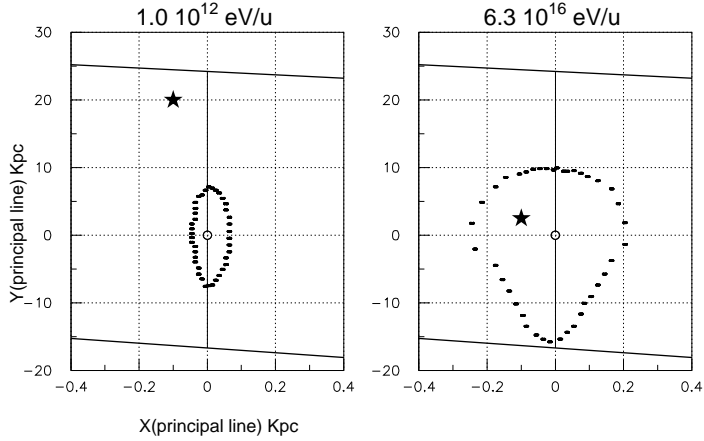}
\vspace{-0.8cm}
\caption[h]{ Contour plots containing $90 \%$ of iron sources
at the energy of  $10^{12}$ and $ 6.3 \times 10^{16}$ $eV$/$u$ in the 
spiral reference frame. These contours
refer to the same iron samples  shown in figure 2 and are called $galactic$ $basins$.
The energy of the right contour is between the $knee$ and $ankle$ energy region.
Galactic basins enlarge with increasing energy,  going from oval
to  ram-head shapes. This enlargement causes an increase of the cosmic ray
intensity at Earth.}
\end{figure}
%%%%%%%%%%%%%%%%%%%%%%%%%%%%%%%%%%%%%%%%%%%%%%%%%%%%%%%%%%%%%%%%

\par Figure 3  displays a contour plot encompassing $90 \%$ of
the sources emitting cosmic rays intercepting the local galactic zone.
The contours in figure 3 refer to the same iron source distributions in figure 2
at the two particular (arbitrary) energies of $5.6  \times 10^{14}$ and
$3.5 \times 10^{18}$ $eV$.
These particular source patterns delimited by the dotted curve surrounding the 
recording instrument are  called here $galactic$ $basins$. They are only a subset
of the total number of sources present in the disk.
In this paper the recording instrument is the Earth, also referred to as
$local$ $galactic$ $zone$ or $solar$ $cavity$
(the black dot in figure 1).

Let us emphasize some features of the galactic basins in figure 3. 
The star outside the iron basin of $10^{12}$ $eV$/$u$ represents a source.
Since this source does not belong to the  basin it implies 
that, on average, cosmic rays emitted from it
do not reach the Earth but, more likely, they will abandon the
disk, or eventually, they will suffer nuclear collisions in the disk.
The star inside the iron basin of $6.3 \times 10^{16}$ $eV$/$u$, on the 
contrary, emanates cosmic rays 
that, with a probability higher than $90 \%$,  will arrive at Earth. Therefore,
this source  belongs to the basin.

\par Generally, cosmic ray intensity registered by a terrestrial
instrument has to diminish as the energy increases, above $ 10^{14}$ $eV$,  
if the number of sources feeding the solar cavity diminishes
 with increasing energy. There are two competing processes, 
referred to as $\alpha$ and  $\beta$, controlling the number of sources powering 
the flux at Earth: one  ($\alpha$) causing  
an increase of the flux,
another ($\beta$) causing a decrease. The basin size increases 
with energy as clearly displayed in figure 3 where the energy varies from  
 $10^{12}$ to  $ 6.3 \times 10^{16}$ $eV$/$u$
 (see also figure 8, Paper III).
As a consequence,  the number of sources increases with the energy, resulting in
a higher flux (process $\alpha$). At the same time, as the energy 
increases,  the average  
length of cosmic ray trajectories in the Galaxy diminishes, 
implying a  decrease of the flux in the local zone (process $\beta$).  
Since the serpentine profile
of the trajectories at low energy (see figure 3, Paper I)
becomes less and
 less intricate with increasing energy (reaching the rectilinear propagation at 
 $ 4 \times 10^{18}$ $eV$ for helium),  the probability
of intercepting the local galactic zone decreases at higher energies. 
In fact, intricate trajectories, under the same physical conditions, 
 yield higher fluxes than rectilinear trajectories, as 
intuition suggests and detailed calculations demonstrate.

 \par The process $\alpha$,  intrinsically, would have yielded  a large 
flux enhancement with increasing energy,  at Earth, in the interval 
$10^{12}$-$10^{17}$ $eV$, if the thickness 
of the Galaxy would have been much 
larger (say, exaggerating, 1000 pc) than its real value of hundreds $pc$. 
The small disk thickness boosts the escape probability against the gain
in the number of sources (process $\alpha$), intrinsic to 
the basin enlargement, as the energy increases. 
Quantitatively, there is a very small flux enhancement at Earth, as the 
energy increases 
between  $6 \times 10^{12}$ and $2 \times 10^{15}$ $eV$ ($5.8 \%$  for He),  
because of the finite disk thickness which erodes the full enhancement effect of the 
process $\alpha$.

\par A third 
effect, referred to as 
process $\gamma$, is caused by the characteristic trend of the inelastic
nuclear cross sections, which rise with energy,  making
the average trajectory length in the disk shorter, compared
to the average length evaluated by constant cross sections. 
And this effect continuously and permanently influences both processes $\alpha$ 
and $\beta$, discussed above and conceived at constant cross section, for 
simplicity. The  only way to quantify  the effect of rising  cross sections 
is to calculate the same physical quantities (intensities, energy spectra etc.)
with constant cross sections,  chosen at an arbitrary energy ($10^{11}$ $eV$) 
and to compare the differences.  This will be done in the next Section 4.
The relative importance of the
processes $\alpha$, $\beta$ and $\gamma$ at different energies
creates the knees of the individual ions as
will be quantitatively clarified in the next Section.

%%%%%%%%%%%%%%%%%%%%%%%%%%%%%%%%%%%%%%%%%%%%%%%%%%%%%%%%%%%%%%%%
\begin{figure}[t!]  %%% FIGURE 4 %%%
\epsfysize=8cm
\hspace{1.8cm}
%%%%%\epsfbox{fig15-zloc-sigstandard-a-20060317-he-frfr.eps} \vspace{0.3cm}
%%%%%\epsfbox{fig15-zloc-sigstandard-a-20060317-he-fr.eps} \vspace{0.3cm}
\epsfbox{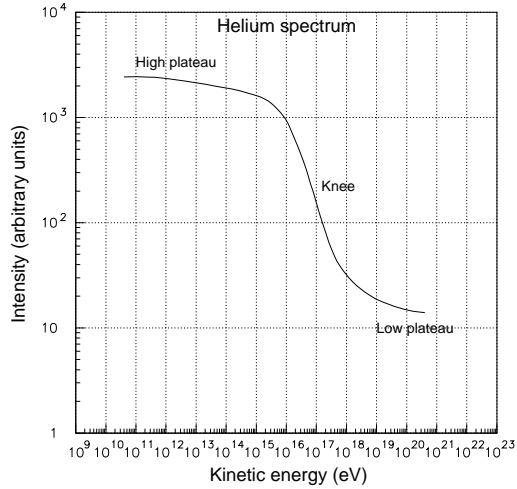}
\vspace{0.3cm}
\vspace{-1.0cm}
\caption[h]{ Number of cosmic rays (helium) reaching the local galactic zone $n_g$ 
versus energy with an arbitrary normalization.
The structure of the curve presents a high plateau, a steep descent and a low plateau.}
\end{figure}
%%%%%%%%%%%%%%%%%%%%%%%%%%%%%%%%%%%%%%%%%%%%%%%%%%%%%%%%%%%%%%%%

\section{ Nuclear collision rate in the whole disk}

In the previous Section 3 using the notion of galactic basin and a variety of arguments, 
a potential fall or rise of the cosmic-ray intensity versus energy has been anticipated
at very high energy.  
Galactic basins specify
the clustering of the sources related to the position of a given 
recording instrument. 
Galactic basins are just geometrical figures in the disk volume linking the sources
 and the recording instrument. 
Generally, the particle flux
is not linear with energy.
For example, the volume of a galactic basin might be proportional 
to the flux on a recording instrument, only if a uniform 
distribution of sources populates the region around the instrument, 
and eventually, other pertinent 
phenomena are approximately constant with varying energy.

\par Instead of dissecting the cosmic-ray flux  in partial contributions
 in order to comprehend the origin of the knees in a coherent rational scheme
as made in the previous section, 
it is also useful and complementary to calculate
the number of cosmic rays reaching the local galactic zone, hereafter referred to
 as $n_g$. This quantity  
versus energy is shown in figure 4 for helium. Note that  
a cosmic ion
emitted from a source with an energy $E_s$ will reach the local galactic 
zone with the same energy $E_s$,  since at these high energies ionization energy
 losses are negligible.

The quantity $n_g$ is approximately proportional to the energy spectrum
of the cosmic rays when the spectral index at the sources is taken into account.
 As apparent from figure 4
the spectrum presents three different regions: (1) a high plateau; (2) a steep descent;
(3) a low plateau. Figure 5  reports $n_g$ versus energy, for six ions, 
representative of important nuclear species.
%%%%%%%%%%%%%%%%%%%%%%%%%%%%%%%%%%%%%%%%%%%%%%%%%%%%%%%%%%%%%%%%
\begin{figure}[t!]  %%% FIGURE 5 %%%
\epsfysize=8cm \hspace{1.5cm}
%%%%%\epsfbox{fig15-zloc-sigstandard-r-20060317-noyau-interpola.eps} \vspace{0.3cm}
\epsfbox{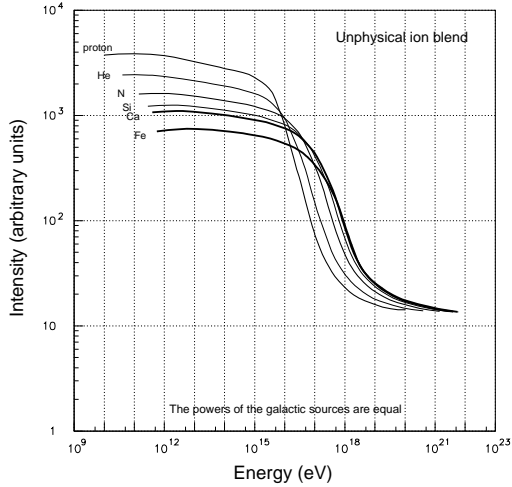}
\vspace{0.3cm}
\vspace{-1.0cm}
\caption[h]{ Energy spectra of six nuclear species obtained by ideally
assuming equal source powers of any galactic ions (referred to as
unphysical ion blend). This unphysical condition visualizes the regularity and
 smoothness of the 
spectra for different nuclear species. The differences in the intensities 
 at a given energy
are due to the nuclear cross sections and to the different grammages experienced
by cosmic ions while propagating in the disc.}
\end{figure}
%%%%%%%%%%%%%%%%%%%%%%%%%%%%%%%%%%%%%%%%%%%%%%%%%%%%%%%%%%%%%%%%
%%%%%%%%%%%%%%%%%%%%%%%%%%%%%%%%%%%%%%%%%%%%%%%%%%%%%%%%%%%%%%%%
\begin{figure}[b!]  %%% FIGURE 6 %%%
\epsfysize=7cm
\hspace{2.3cm}
%%%%%\epsfbox{fig10-norm-bassin-haint-sigconstant-a.eps} \vspace{0.3cm}
\epsfbox{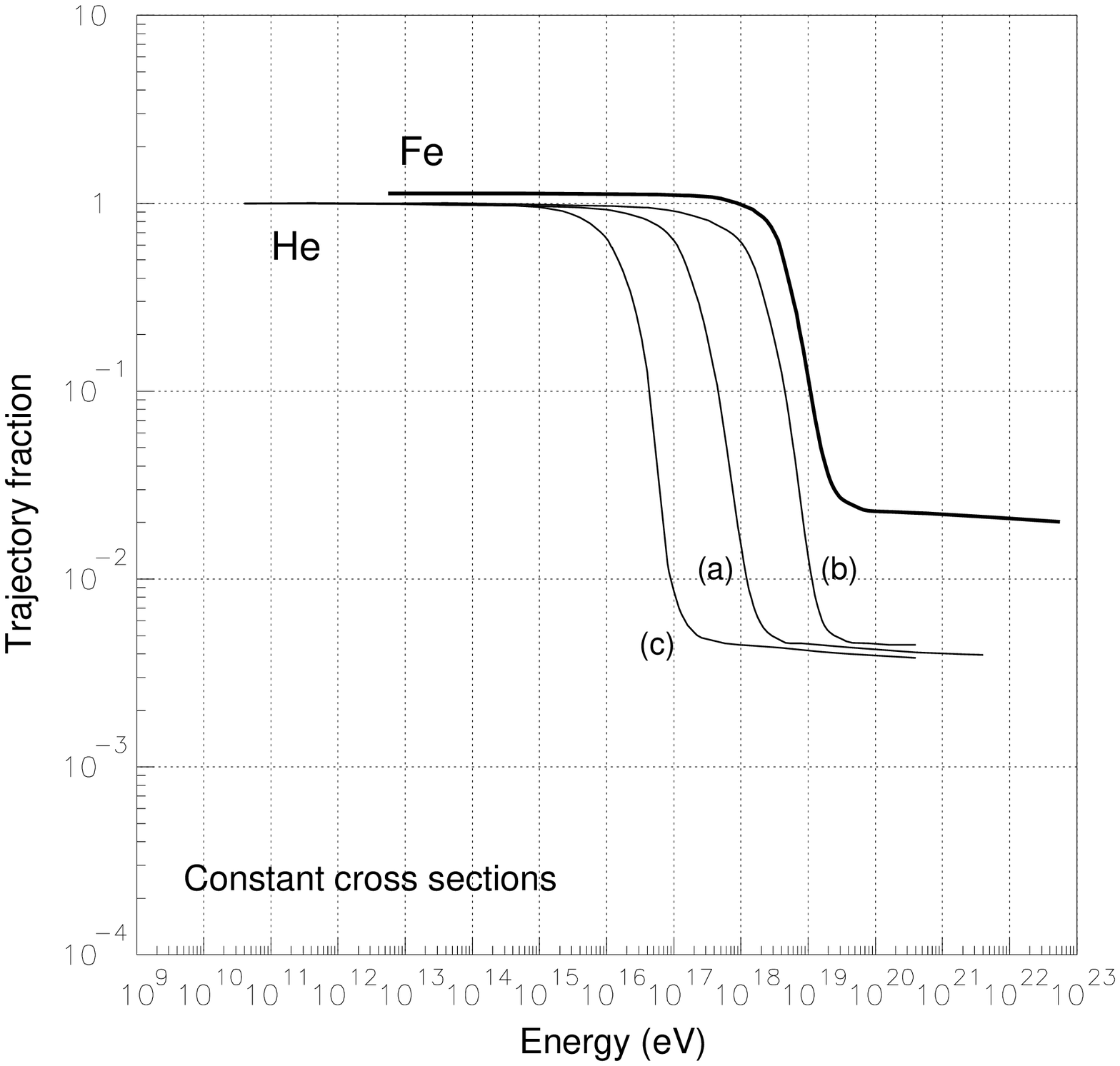}
\vspace{0.3cm}
\vspace{-0.8cm}
\caption[h]{ Effect of the magnetic field strength on the 
nuclear collision rate of helium and iron occurring in the disk. The 
correct magnetic field strength refers to curve (a) while in curve (b) 
the strength is artificially enhanced by a factor 10, and  
curve (c) decreased by a factor 10.
Helium-proton and iron-proton nuclear cross sections of 102 mb and  718 
mb, respectively, at the arbitrary energy of 100 $GeV$, are adopted 
in this particular calculation; they are artificially taken constant 
in the energy interval $10^{10}$-  
$10^{21}$ $eV$, just to single out and quantify  the difference on the flux
 against 
(real) rising cross sections. Trajectories fractions are normalized
to 1 for helium, at the energy of $10^{12}$ $eV$. }
\end{figure}
%%%%%%%%%%%%%%%%%%%%%%%%%%%%%%%%%%%%%%%%%%%%%%%%%%%%%%%%%%%%%%%%

\par In order to appreciate the forms of the spectra 
given in figure 4 and 5 it is useful to determine the number of cosmic rays 
suffering nuclear collisions in the whole disk for an arbitrary initial sample (several
million fully reconstructed cosmic-ray trajectories have been used for the
results in figure 5). Cosmic rays either escape from the disk or are 
destroyed by nuclear collisions and 
the probabilities for the two processes are intimately related.
Any given sample of trajectories with sources rooted
 in the
disk volume must satisfy the equation: $f_N$ + $f_E$ = 1, where $f_N$ is the fraction
of trajectories undergoing nuclear collisions and $f_E$ that escaping from the disk.
Note that, at high energy,  the fraction of trajectories extinguished by ionization 
energy losses, $f_I$, is zero and consequently omitted from the above equation.

\par The number of nuclear collisions versus energy in the whole disk is shown in 
figure 6.
Curves (a),  (b) and  (c) refer to helium while the solid black curve
to iron.  Curve (a) is obtained by the set of parameters mentioned in Section 2
(real, observational  parameters) while, for  curve (b),  the field strength
 is artificially multiplied by a factor 10, and  for curve (c), reduced by a factor 10.
Of course, the iron spectrum in figure 6 also splits into three similar curves 
which, for simplicity, are not shown.

\par The effect of the magnetic field is to bend and invert particle trajectories
 during ion propagation in the disk.  When this process is at full
 regime,  the matter thickness encountered by cosmic rays (the grammage)  is constant
and the resulting  number of nuclear collisions is also constant, if nuclear 
cross sections are constant. The high plateau
between  $ 10^{12}$  and $ 10^{16}$ $eV$ for helium
and between $ 10^{13}$ and $ 10^{18}$ $eV$ for iron exhibits  this condition.
When the bending power of the magnetic field, around $10^{16}$ $eV$ for helium, becomes 
inefficient (or less efficient than that of the magnetic field configuration 
operating at  lower energies) trajectories become less and less twisted, 
and consequently, the
probability of escaping from the disk grows. Equivalently
stated,  the number (or the fraction)  of nuclear collisions decreases
 with increasing energy, and since   $f_E$ = $1$ - $f_N$,  the rate of particle escape 
increases.
Hence, cosmic ray intensity in the disk abruptly reduces 
and a steep fall  appears.  But the descent
cannot continue indefinitely. Once the magnetic field has completely lost its 
bending power, 
so that cosmic  ions propagate in straight-line segments,  the matter thickness attains 
its minimum value; in this condition, the low plateau with a constant, minimum number of 
nuclear collisions is attained.  The  
separation of the energy spectra (a), (b) and (c) in figure 6 vividly and beautifully
manifests how the particular bend energy in the spectrum is controlled by the 
average magnetic field 
in the Galaxy. Table 1 firmly attests the experimental basis 
of the average field strengths in spiral galaxies which,
beyond any doubt, fix and delimit along the energy axis, the realm of 
the rectilinear propagation.

%%%%%% Table 2 %%%%%%%%%%%%%%%%%%%%%%%%%%%%%%%%%%%%%%%%%%%%%%%%%%%%%%%%%%%%
\begin{table}[t!]
\begin{center}
\caption{ Relative abundances of cosmic ions and spectral indices
at the arbitrary energy of $2 \times 10^{15}$ $eV$.
The ion groupings CNO, Ne-S and Fe(17-26) derive from the 
traditional data analysis in many experiments. The computed spectral
index of the complete spectrum is denoted by $\gamma$.} 
\bigskip
\begin{tabular}{lllllll}

\hline
            & Blend 1  &          & Blend 2   &          & Blend 3  &          \\
\hline
            & Proton   &          & Proton    &          & Helium   &          \\
            & abundant &          & superabundant&       & abundant &          \\
\hline
Ion         & Comp.    & Index    & Comp.     & Index    & Comp.    & Index    \\
\hline
H           & 32.8$\%$ & 2.72     & 37.9$\%$  & 2.74     & 19.6$\%$ & 2.72     \\
He          & 29.7$\%$ & 2.72     & 27.4$\%$  & 2.72     & 35.5$\%$ & 2.72     \\
CNO         & 11.7$\%$ & 2.65     & 10.8$\%$  & 2.65     & 14.0$\%$ & 2.65     \\
Ne-S        & 10.0$\%$ & 2.65     &  9.3$\%$  & 2.65     & 12.0$\%$ & 2.65     \\
Fe(17-26)   & 15.0$\%$ & 2.60     & 13.8$\%$  & 2.60     & 17.8$\%$ & 2.50     \\
Ca          &  0.8$\%$ & 2.60     &  0.8$\%$  & 2.60     &  0.1$\%$ & 2.60     \\
\hline
$\gamma$    &          & 3.05     &           & 3.06     &          & 3.00     \\
\hline
\end{tabular}
\end{center}
\end{table}
%%%%%%%%%%%%%%%%%%%%%%%%%%%%%%%%%%%%%%%%%%%%%%%%%%%%%%%%%%%%%%%%%%

\par  Detailed calculations indicate that the
transition from the results shown in figure 6 to those reported in figure
4 and 5 (energy spectra to be compared with the experimental data) is
accounted for by the effect of rising cross sections and the position of the 
local zone in the Galaxy.
The similarity of the curves demonstrates that the break of the high plateau 
caused by the magnetic field is a general phenomenon occurring both in the local zone 
(a sphere of 50 $pc$ in diameter in this particular calculation) and in the whole disc.  
The method of calculation used for the results given in figure 6 is described
elsewhere (see Section 4, Paper II).

%%%%%%%%%%%%%%%%%%%%%%%%%%%%%%%%%%%%%%%%%%%%%%%%%%%%%%%%%%%%%%%%
\begin{figure}[h!]  %%% FIGURE 7 %%%
\epsfysize=9cm \hspace{1.5cm}
%%%%%\epsfbox{spectre-genou-schema.eps}
\epsfbox{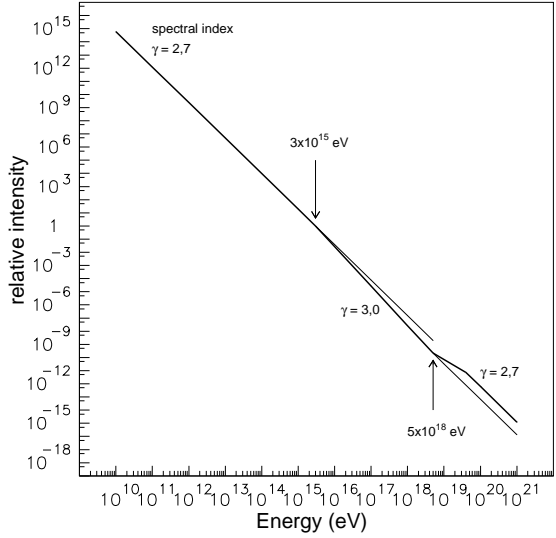}
%%%%%\vspace{0.3cm}
\vspace{-0.9cm}
\caption[h]{ Universal energy spectrum of the cosmic radiation
measured at Earth in the interval $10^{10}$-  
$10^{21}$ $eV$ with the relevant spectral indices, and with knee and ankle energies 
indicated. The thin line above $3 \times 10^{15}$ $eV$ is the spectrum extrapolation 
with an index of 2.7 and that above $5 \times 10^{18}$ $eV$ is the extrapolation 
with an index of 3.0. The difference between measured and extrapolated spectra
suggests that the $knee$ and the $ankle$ are small perturbations. 
The approximate equality of the spectral index over the entire
energy range lends credence to the existence of a universal acceleration 
engine operating inside and outside galaxies 
as argued elsewhere (Codino, 2005). }
\end{figure}
%%%%%%%%%%%%%%%%%%%%%%%%%%%%%%%%%%%%%%%%%%%%%%%%%%%%%%%%%%%%%%%%

%%%%%%%%%%%%%%%%%%%%%%%%%%%%%%%%%%%%%%%%%%%%%%%%%%%%%%%%%%%%%%%%
\begin{figure}[t!]  %%% FIGURE 8 %%%
\epsfysize=6.7cm \hspace{2.0cm}
%%%%%\epsfbox{fig20a-experimentale-proton-200603-he2608-bis-perfect.eps}
\epsfbox{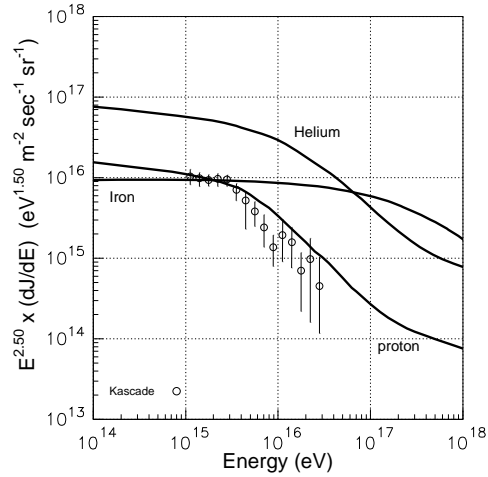}
%%%%%\vspace{0.3cm}
\vspace{-0.5cm}
\caption[h]{ Proton energy spectrum compared with the experimental
data measured by the Kascade Collaboration elaborated by the SIBYLL algorithm.
The proton spectrum measured by the Tibet AS$\gamma$ Collaboration 
(Amenomori, 2006) seems
in disagreement with the present evaluation and incompatible with the Kascade 
outcomes (the
 bend energy and spectrum form as well).}
\end{figure}
%%%%%%%%%%%%%%%%%%%%%%%%%%%%%%%%%%%%%%%%%%%%%%%%%%%%%%%%%%%%%%%%
%%%%%%%%%%%%%%%%%%%%%%%%%%%%%%%%%%%%%%%%%%%%%%%%%%%%%%%%%%%%%%%%
\begin{figure}[t!]  %%% FIGURE 9 %%%
\epsfysize=6.7cm \hspace{2.0cm}
%%%%%\epsfbox{fig20c-experimentale-eastop-etot-vent-aa.eps}
\epsfbox{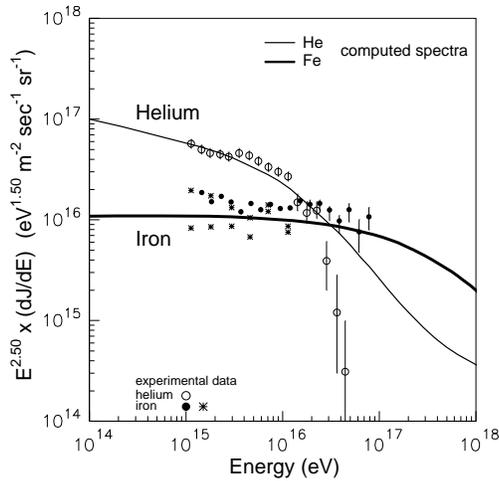}
%%%%%\vspace{0.3cm}
\vspace{-0.5cm}
\caption[h]{  Computed helium and iron energy spectra compared with the 
experimental data. Open dots (He) and full dots (Fe) from the Kascade
Collaboration; stars (Fe) from the Eas-top Collaboration indicate the maximum
and minimum intensity at the specified energy.}
\end{figure}
%%%%%%%%%%%%%%%%%%%%%%%%%%%%%%%%%%%%%%%%%%%%%%%%%%%%%%%%%%%%%%%%

\section{Computed and measured proton, helium and iron energy spectra}

The comparison of the computed energy spectra with the measured ones,
in the energy band $10^{15}$-$ 10^{19}$ $eV$, although
simple and feasible,  requires the spectral 
indices and chemical composition of the individual  cosmic ions at the 
cosmic-ray sources.
Figure 7 shows the differential energy
spectrum of the cosmic radiation in the interval  $10^{10}$-$10^{21}$ $eV$ 
with some characteristics.
In spite of the two significant deviations at $ 3 \times 10^{15}$ and  
$ 5 \times 10^{18}$ $eV$, the spectrum is 
remarkably linear in the logarithmic scales of intensity
 and energy. The subsequent analysis of the knees takes advantage of this linearity
by assuming constant spectral indices for any 
ion in the whole  energy range $10^{10}$-$10^{21}$ $eV$. 
This ultrasimplified hypothesis is indeed supported by many experimental data
(on the absence of a spectral break at low energy see, for example,  Cherry, 1999).
Notice however, that any  
other hypothesis, in the form of two or more values of the
spectral indices in the energy range $10^{10}$-$10^{21}$ $eV$ 
would yield an even better agreement 
in the comparison with the experimental data.

\par Traditionally, leaky box models of cosmic rays subdivide the
spectral index (always close to 2.7) of a nuclear species observed at Earth   
as the sum of two parts:  one due to the sources (typically 2.2) and the rest  due to 
the propagation. 
The logical necessity for this splitting 
remains, to date, largely undemonstrated and the observational evidence contradictory,
as it results from the comparison of the spectral index at the sources
extracted from the analysis of the
anisotropy of  the arrival direction of high energy cosmic 
rays (Hillas, 1984) and that extracted from the
 $\gamma$-ray spectrum recently measured by the HESS Collaboration
(Aharonian, 2006).
Accordingly, the hypothesis for this 
splitting is not 
retained. An empirical approach
is adopted here,  where spectral indices of the individual ions in the whole disk are
assumed to be those measured at Earth. $A$ $posteriori$ this crude approximation
is more than adequate for the explanation of the knees. 

%%%%%%%%%%%%%%%%%%%%%%%%%%%%%%%%%%%%%%%%%%%%%%%%%%%%%%%%%%%%%%%%
\begin{figure}[t!]  %%% FIGURE 10 %%%
\epsfysize=8cm \hspace{1.6cm}
%%%%%\epsfbox{fig20a-experimentale-proton-200603-he2720-noyaux.eps}
\epsfbox{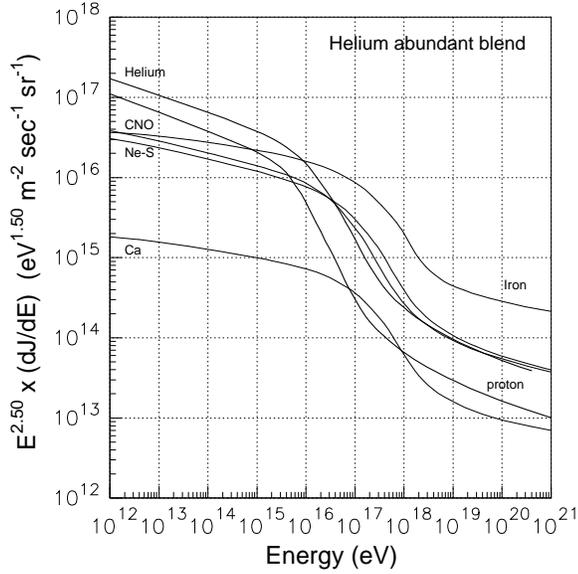}
%%%%%\vspace{0.3cm}
\vspace{-0.5cm}
\caption[h]{  Computed energy spectra of individual ions according
to the elemental abundances and spectral indices given in Table 2 referred
to as blend 3. 
These curves are just one of the many examples of how the energy spectra 
given in figure 5 may be transformed by a particular ion blend. }
\end{figure}
%%%%%%%%%%%%%%%%%%%%%%%%%%%%%%%%%%%%%%%%%%%%%%%%%%%%%%%%%%%%%%%%
\par In Table 2 are given three elemental compositions and related spectral indices
(hereafter called ion blends) at the arbitrary energy of $2 \times 10^{15}$ $eV$
 where the 
equalization of the computed energy spectra to the experimental ones is imposed.

%%%%%%%%%%%%%%%%%%%%%%%%%%%%%%%%%%%%%%%%%%%%%%%%%%%%%%%%%%%%%%%%
\begin{figure}[t!]  %%% FIGURE 11 %%%
\epsfysize=8cm \hspace{1.6cm}
%%%%%\epsfbox{fig20a-experimentale-extraga-200603-he2720-haverah-tibet-3p-extra.eps}
%%%%%\epsfbox{fig20a-experimentale-extraga-200603-he2720-haverah-tibet-3p-extra-catania.eps}
\epsfbox{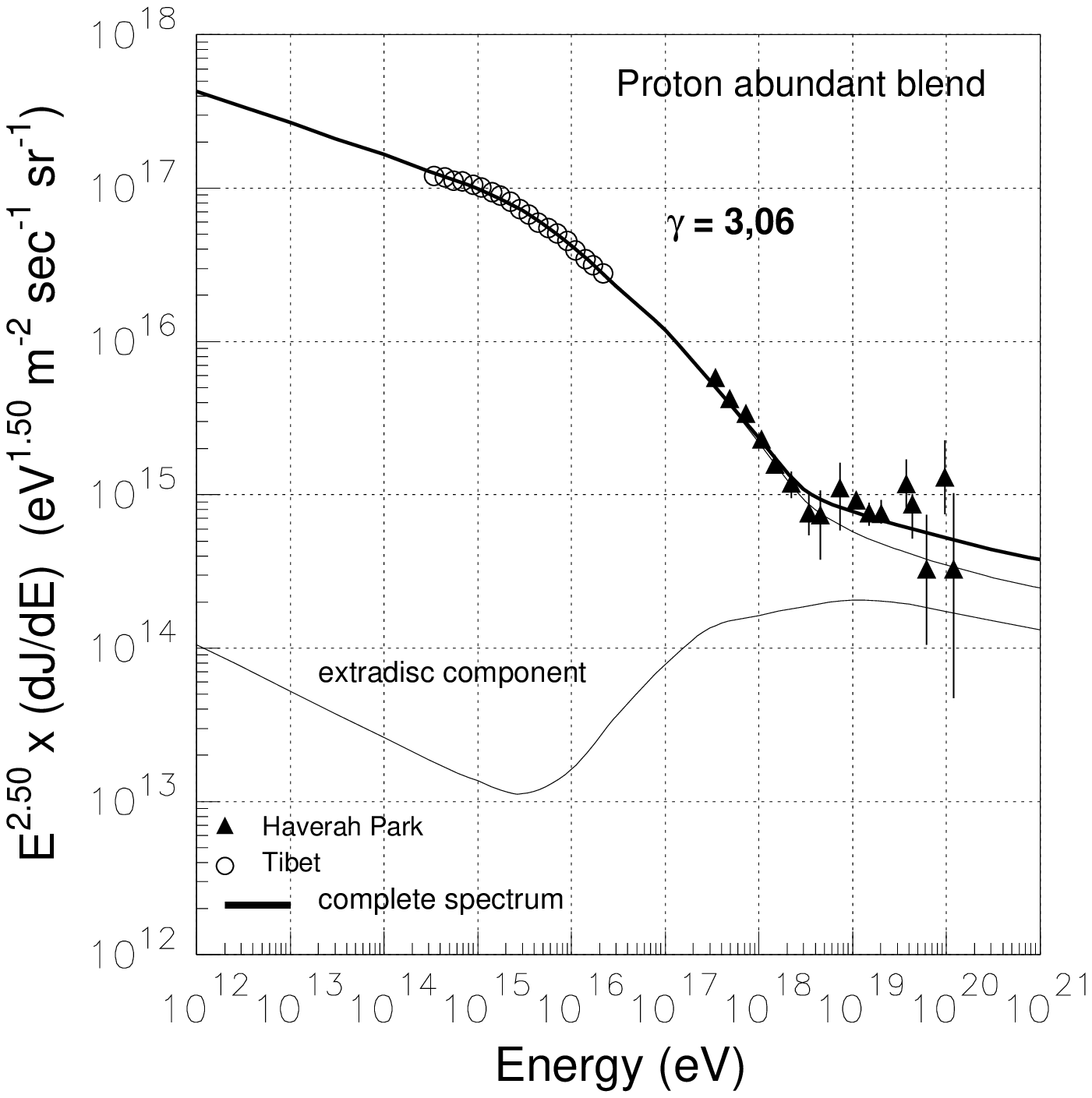}
%%%%%\vspace{0.3cm}
\vspace{-0.5cm}
%%%%%\caption[h]{  Complete energy spectrum in the energy band $10^{10}$-$ 10^{21}$
%%%%% $eV$ computed with the ion blend 2 given in Table 2 and its comparison 
%%%%%with the experimental data of the Tibet and Haverah 
%%%%%Park experiments. }
\caption[h]{  Complete energy spectrum in the range $10^{12}$-$ 10^{21}$
 $eV$ with the ion blend 2 of Table 2 and its comparison 
with the Tibet and Haverah Park data. }
\end{figure}
%%%%%%%%%%%%%%%%%%%%%%%%%%%%%%%%%%%%%%%%%%%%%%%%%%%%%%%%%%%%%%%%

%%%%%%%%%%%%%%%%%%%%%%%%%%%%%%%%%%%%%%%%%%%%%%%%%%%%%%%%%%%%%%%%
\begin{figure}[t!]  %%% FIGURE 12 %%%
\epsfysize=8cm \hspace{1.6cm}
%%%%%\epsfbox{fig20a-experimentale-extraga-200603-he2720-akeno-agasa-3p-extra.eps}
%%%%%\epsfbox{fig20a-experimentale-extraga-200603-he2720-akeno-agasa-3p-extra-catania.eps}
\epsfbox{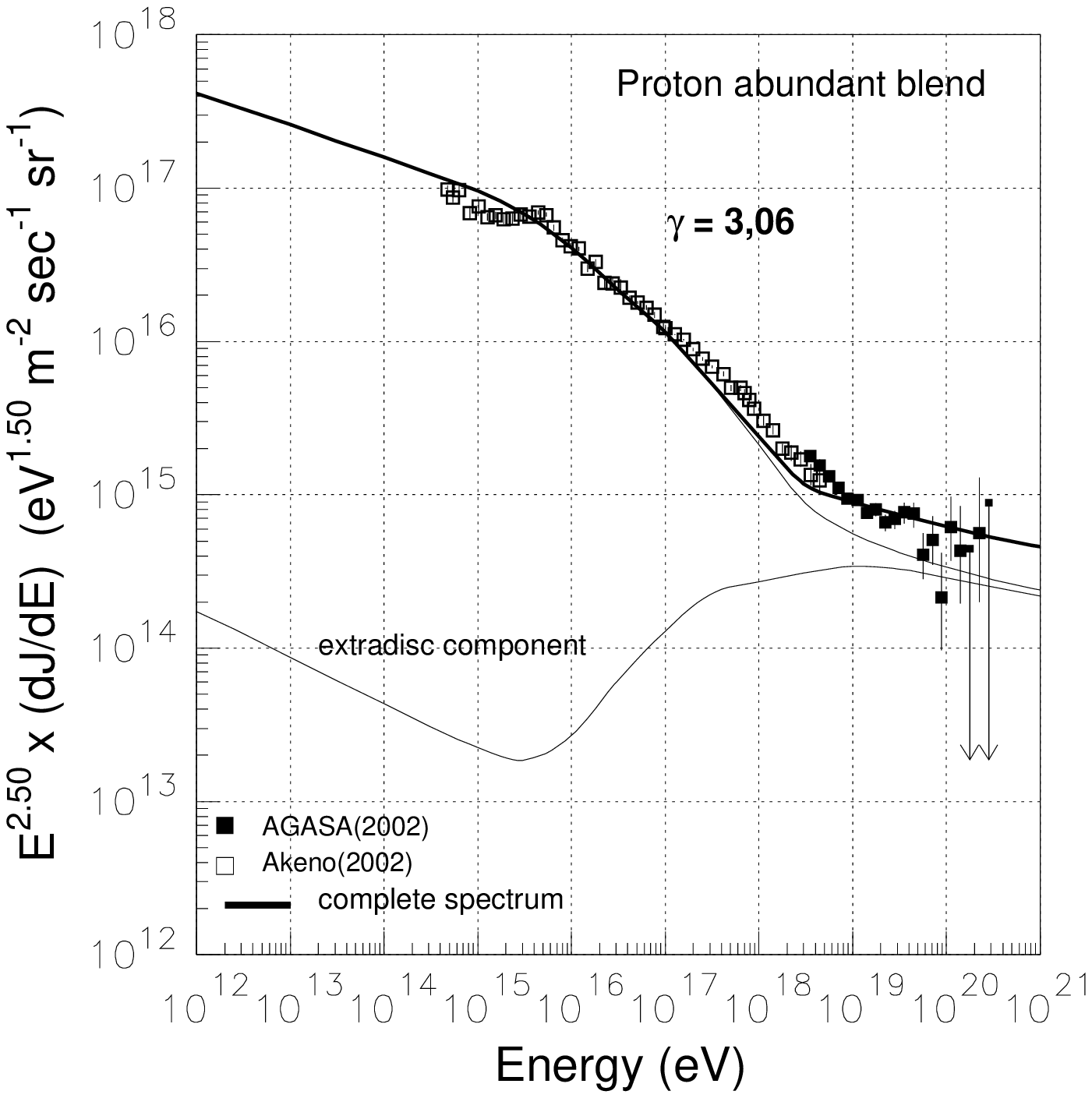}
%%%%% \vspace{0.3cm}
\vspace{-0.5cm}
\caption[h]{  Complete energy spectrum in the range $10^{12}$-$ 10^{21}$
 $eV$ with the ion blend 2 of Table 2 and its comparison with 
the Akeno and Agasa data. }
%%%%%\caption[h]{  Complete energy spectrum in the energy band $10^{10}$-$ 10^{21}$
%%%%% $eV$ with the ion blend 2 given in Table 2 and its comparison with 
%%%%%the experimental data of the Akeno and Agasa 
%%%%% experiments. }
\end{figure}
%%%%%%%%%%%%%%%%%%%%%%%%%%%%%%%%%%%%%%%%%%%%%%%%%%%%%%%%%%%%%%%%

Let us anticipate 
here that by altering the values of
the ion blends, within plausible limits bound to  measurements,
the agreement between computed and measured energy spectra
substantially persists. Notice, on the contrary,
that the 
spectral indices become critical, for the accord with the data, at energy regions 
quite remote  from the knee energy band (e.g. at the ankle energy region, see
subsequent figures 11 and 12).  
Figure 8 shows the proton energy spectrum with a spectral index of 2.6,
based on the Atic experiment, at very low energy (Adams et al., 2005), normalized 
to the flux  of $0.97 \times 10^{-22}$ particles/$eV$ $m^2$$sr$ $sec$,  
measured by the Kascade Collaboration (Roth et al., 2003)
 at the energy of $ 2 \times 10^{15}$ $eV$.
The computed spectrum conforms fairly well to the measurements; 
taking a soft spectral index (for example,  2.72) instead of a hard one,     
the agreement with the experimental data becomes excellent. The energy
spectrum  obtained by the Sibyll algorithm in Kascade gives a similar accord,
 since the
resulting spectra almost shift in parallel manner in intensity. Note that
 the mentioned shift
in intensity is absorbed in the normalization, so preserving the accord. 
\par In figure 9 are given the computed helium and iron spectra (thick lines) 
compared with the experimental data of Kascade and Eas-top Collaborations
 (Navarra et al., 2003).
In this case, the helium spectral index of 2.72 has been used and the traditional
value of 2.5 for iron (Mueller, 1989), as measured at low energy. 
The helium-to-iron flux ratio of 3.75 at the normalization energy 
of $ 2 \times 10^{15}$ $eV$ has been imposed using the data from the
Kascade Collaboration (Roth et al., 2003). 
Comparing the computed helium
spectra in figures 8 and 9 (thick lines in both figures), which have different
 spectral indices of 
2.6 and 2.72,  a visual quantification of the general  effect of the indices
on the intrinsic spectra given in figure 5,  may be appreciated. 
The global agreement with the measured H, He and Fe spectra is more than 
satisfactory.

\section{The complete spectrum of the cosmic radiation between the knee and the ankle}

\par The complete spectrum of the cosmic radiation is the sum of the 
 partial spectra of the individual ions. Figure 10 shows an example of
 partial spectra for six ions.
The indices and elemental abundances of these energy spectra are those 
reported in Table 2, blend 3.
The regular, smooth spectra displayed in figure 5, which have (artificial) equal
abundances at the sources, 
transform into those shown 
in figure 10, which are scattered in intensity, due to the uneven chemical 
composition and differing spectral indices, ranging from 2.60 to 2.72. 

\par In figures 11 and 12 
the complete energy spectra for the ion blend 2, in the interval 
$10^{10}$-$10^{21}$ $eV$,
are given,  along with the data of some groups : 
Tibet and Haverah Park experiments in figure 11
(Amenomori, 1996; Lawrence, 1990) Akeno and
 Agasa experiments in figure 12 
(Hara, 1979; Nagano, 1992).
The computed spectrum is required to agree with the experimental data
only at one energy point, at the arbitrary energy of $ 1.0  \times 10^{16}$ $eV$ 
and no other constraint is imposed. For sake of completeness, at the energy of 
$ 1.0  \times 10^{16}$ $eV$, the intensity ratio of the complete spectra 
between Tibet and Kascade experiments is taken 1.4
while that between Akeno and Kascade 0.65. 
The overall accord is fairly good, and in particular,
the spectral index $\gamma$ of 3.06 for this ion blend 2 
in the energy interval $ 6.0  \times 10^{15}$-$10^{17}$ $eV$ is 
in excellent agreement
with the data of the quoted experiments and many others.
The other ion blends 1 and 3 give similar results for $\gamma$,
3.05 and 3.0, respectively,
 as reported in Table 2.

\par In the present investigation of the energy spectrum between the knee and the ankle,
the traditional hypothesis (see, for example, Rossi B., 1964) is admitted that 
a fraction of the sources powering
the flux at Earth might  not be located in the disc volume. Let us call
these sources extradisc rather than extragalactic since there are claims
of sources placed in the halo (Hillas, 1998; Dar, 2001; Plaga, 2003). 
In the next  Section it is
demonstrated that the expected extradisc flux, if any, has to peak between the energies 
of $ 3 \times 10^{18}$ and $ 5 \times 10^{19}$ $eV$, namely, the minimum (for protons) 
 and  maximum energies (for Fe) attesting  the rectilinear propagation of cosmic rays
in the Milky Way.
Incorporating this anticipated result here,
the extradisc component ($I_e$)  is normalized at the energy of
$ 1.0  \times 10^{19}$ $eV$. The normalization requires that the sum 
of the extradisc 
and galactic intensity at Earth ($I_g$) equals the observed value of   
$2.5 \times 10^{-33}$ particles/$eV$ $m^2$$sr$  $sec$ as measured by the Haverah Park 
experiment (figure 11) or by $2.25 \times 10^{-33}$ particles/$eV$ $m^2$$sr$  $sec$ 
as measured by Agasa Collaboration (figure 12). 
It is found that
($I_g$+$I_e$)/$I_e$ is about 3.6 in figure 11 and ($I_g$+$I_e$)/$I_e$ about  2.5 
in figure 12.
\par The energy spectra of the extradisk component are determined
 by taking into account the modification  
of the energy spectra and ion blends  (with respect to those given in Table 2) 
experienced by the galactic cosmic rays overflowing from the Galaxy and 
the analogous modifications suffered by the extragalactic cosmic 
rays penetrating the 
Milky Way from its exterior and reaching the solar cavity.  Details of this 
developement will be  given
at the CRIS Catania Conference 2006 (Codino and Plouin, 2006).

%%%%%%%%%%%%%%%%%%%%%%%%%%%%%%%%%%%%%%%%%%%%%%%%%%%%%%%%%%%%%%%%
\begin{figure}[b!]  %%% FIGURE 13 %%%
\vspace{-0.8cm}
\epsfysize=9cm \hspace{2.3cm}
%%%%%\epsfbox{spectre-genou-schema.eps}
%%%%%\epsfbox{confvulcano59xx.eps}
%%%%%\epsfbox{confvulcano-fig-13.eps}
\epsfbox{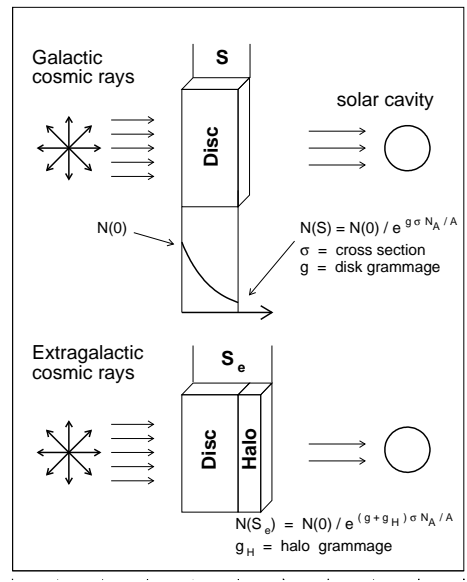}
%%%%%\vspace{0.3cm} 
\vspace{-0.8cm}
\caption[h]{ Schematic illustration of galactic and 
extragalactic sources (sheafs of arrows) contributing to the global flux
at the solar cavity (also Earth). The  slabs of matter represent
 the grammage encountered by cosmic ions
 in the travel to the Earth. The grammage is a global parameter, incorporating 
the structure of the  magnetic field and the 
dimension and the matter density of the traversed medium.}
\end{figure}
%%%%%%%%%%%%%%%%%%%%%%%%%%%%%%%%%%%%%%%%%%%%%%%%%%%%%%%%%%%%%%%%

%%%%%%%%%%%%%%%%%%%%%%%%%%%%%%%%%%%%%%%%%%%%%%%%%%%%%%%%%%%%%%%%
\begin{figure}[h!]  %%% FIGURE 14 %%%
\vspace{-0.8cm}
\epsfysize=8cm \hspace{2.7cm} 
%%%%%\epsfbox{ spectre-genou-schema.eps}
%%%%%\epsfbox{confvulcano60xx.eps}
%%%%%\epsfbox{confvulcano-fig-14.eps}
\epsfbox{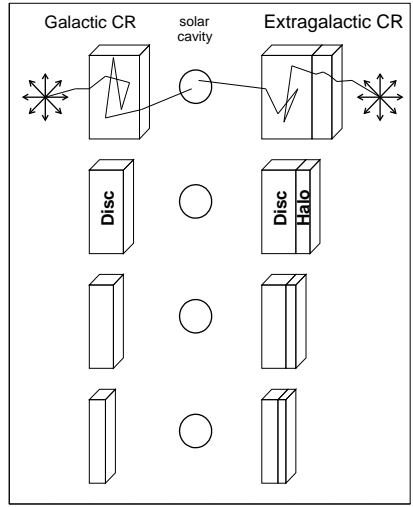}
%%%%%\vspace{0.3cm}
\vspace{-0.8cm}
\caption[h]{ Sequence of barriers of matters (slabs) that
galactic and extradisc cosmic rays should penetrate in order to
contribute to the local flux at Earth. The thicknesses of these barriers
decrease with increasing energy attaining a minimum value at a particular, 
distinctive energy, for a given nuclear species. 
The energy corresponding to this minimum defines the ankle energy. }
\end{figure}
%%%%%%%%%%%%%%%%%%%%%%%%%%%%%%%%%%%%%%%%%%%%%%%%%%%%%%%%%%%%%%%%

\section{ How the knee and ankle energies of  individual
 ions are interconnected}

Consider the extreme simplified condition of galactic and extragalactic cosmic rays
represented in figure 13.
Galactic sources are represented by diverging arrows while the disc and halo
 by slabs of matter with  appropriate thicknesses.
The slab thickness, $s$,  is the average distance experienced
by cosmic rays in the travel from a (fictitious) average disc source to 
the solar cavity (circle) and likewise $s_e$ for extragalactic cosmic rays. 
The disc grammage $g$ (in $g$/$cm^2$) is defined as $m$$n$$L$ where $n$ is the 
average atomic mass 
in the disc, $n$ is the number density of atoms, $L$ the mean trajectory length,
and likewise $g_H$ for the halo grammage.
The corresponding average matter densities experienced by galactic and 
extragalactic ions are : $d$=$g$/$s$ 
and $d_e$=$g_H$/$s_e$,
 respectively. Note that $s$ and $s_e$ are 
constant lengths which disappear from the final formulae. 
\par Galactic sources  emanate cosmic rays
which propagate through the disc and beyond its border. Some of these ions
are destroyed by nuclear collisions in the disk, some
others evade into the halo, and only a tiny fraction arrives to the solar cavity.
Similar partitions of the flux should  occur for the extragalactic component
 (bottom part of figure 13) with the only
difference that the slab thickness, $s_e$, is a bit higher than that experienced
by galactic cosmic rays. This is due to the larger size of the traversed 
medium. 
The matter densities  of these barriers (or the grammages $g$ and $g_H$)  
 certainly vary with the energy as qualitatively shown in figure 14.
Going from low to very high energy, at about 
$10^{18}$ $eV$, the grammage must
reduce to a minimum value which is
attained as cosmic rays propagate
rectilinearly.

%%%%% NOUVELLE PLACE DES TROIS FIGURES 15 16 17

%%%%%%%%%%%%%%%%%%%%%%%%%%%%%%%%%%%%%%%%%%%%%%%%%%%%%%%%%%%%%%%%
\begin{figure}[b!]  %%% FIGURE 15 %%%
\vspace{-1.0cm}
\epsfysize=8cm \hspace{1.8cm}
%%%%%\epsfbox{fig12-sig-gram-200605-cheville-1-he.eps}
\epsfbox{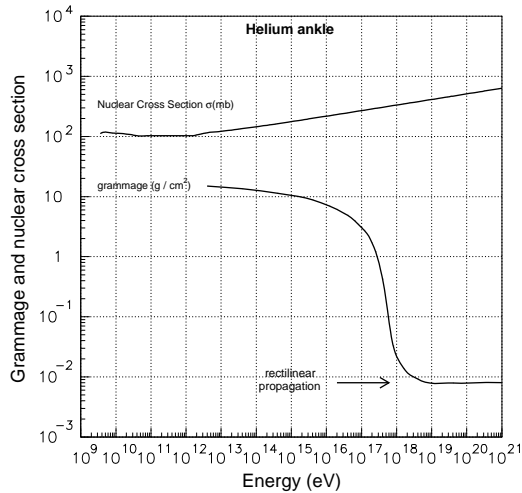}
%%%%%\vspace{0.3cm}
\vspace{-0.8cm}
\caption[h]{ Helium-proton inelastic cross section
versus energy (top curve). Helium grammage versus energy (bottom curve)
in the galactic disc.
The arrow indicates the particular minimum energy of $ 4 \times 10^{18}$  $eV$ 
at which
the grammage attains its asymptotic value. }
\end{figure}
%%%%%%%%%%%%%%%%%%%%%%%%%%%%%%%%%%%%%%%%%%%%%%%%%%%%%%%%%%%%%%%%

%%%%%%%%%%%%%%%%%%%%%%%%%%%%%%%%%%%%%%%%%%%%%%%%%%%%%%%%%%%%%%%%
\begin{figure}[t!]  %%% FIGURE 16 %%%
\vspace{-1.0cm}
\epsfysize=8cm \hspace{2.0cm}
%%%%%\epsfbox{fig12-sig-gram-200605-cheville-2-he.eps}
\epsfbox{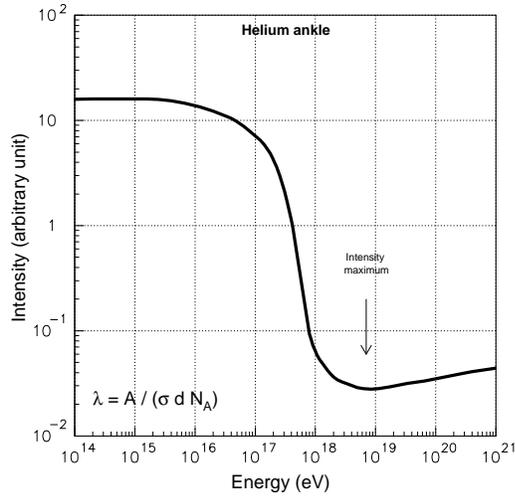}
%%%%%\vspace{0.3cm}
\vspace{-0.8cm}
\caption[h]{ Number of extragalactic cosmic helium ions arriving at Earth 
with an arbitrary normalization. There is a minimum in the barrier thickness
between the extradisc sources and Earth generating the  
intensity maximum at about $7 \times 10^{18}$ $eV$.   
This minimum is  identified with the helium ankle.}
\end{figure}
%%%%%%%%%%%%%%%%%%%%%%%%%%%%%%%%%%%%%%%%%%%%%%%%%%%%%%%%%%%%%%%%
%%%%%%%%%%%%%%%%%%%%%%%%%%%%%%%%%%%%%%%%%%%%%%%%%%%%%%%%%%%%%%%%
\begin{figure}[t!]  %%% FIGURE 17 %%%
\vspace{-1.0cm}
\epsfysize=8cm \hspace{2.0cm}
%%%%%\epsfbox{fig15-zloc-sigstandard-a-echapp-20060317-fe.eps}
\epsfbox{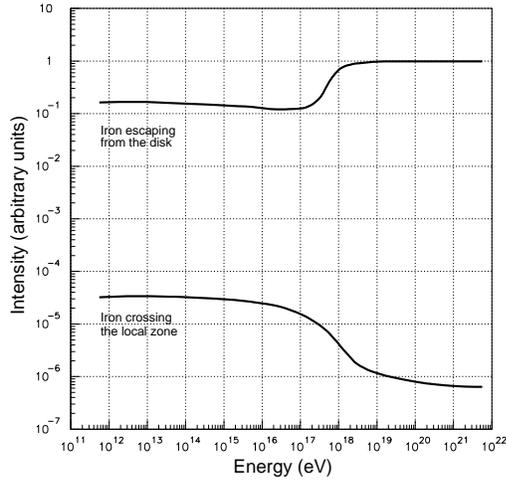}
%%%%%\vspace{0.3cm}
\vspace{-0.9cm}
\caption[h]{ Number of iron nuclei escaping from the disk (top curve)
and those intercepting the local galactic zone
versus energy. An iron knee and antiknee appear as a result of the
different positions in the Galaxy of the two recording instruments.
 The relative intensities of the two curves, based on
$4 \times 10^5$ simulated trajectories, are arbitrarily normalized. 
The computed iron knee (bottom curve) is the same spectrum
shown in figure 9 in accord with the Kascade and Eas-top data. }
\end{figure}
%%%%%%%%%%%%%%%%%%%%%%%%%%%%%%%%%%%%%%%%%%%%%%%%%%%%%%%%%%%%%%%%

\par  The intensity at Earth of the extragalactic component 
is simply determined by the nuclear cross section and the grammage.
As an example, figure 15 displays the helium-proton cross section, which rises 
with energy, and the  helium grammage $g_H$ versus energy.
The average minimum energy at which   cosmic rays penetrate unbent 
the  interstellar medium and the halo, marks 
the onset of the rectililear
propagation and in this condition a minimum grammage is attained. The 
characteristic energy value 
at which this occurs, at $ 4 \times 10^{18}$ $eV$ for helium, 
is due to the average strength of the magnetic field in the Milky Way
and the charge of the ion.
\par  The number of particles surviving the barrier, those arriving at Earth, 
is governed by the attenuation length, $\lambda$, equal to  $A$/$\sigma$$d_e$$N_A$
in the ratio $s_e$/$\lambda$ where $N_A$ is the Avogadro number and 
$A$ the average mass of the galactic atom. 
The product  $s_e$$\sigma$ $d_e$, or more simply $\sigma$$g_H$,  
has necessarily a minimum, shown in figure 16, because of the forms of the 
curves shown in figure 15.   But the  minimum of the
 barrier density  corresponds to the maximum penetration of the barrier, which
in turn, translates in an intensity maximum for the extradisc component.
Repeating this calculation for iron  
the intensity maximum is at $ 5 \times 10^{19}$ $eV$. Therefore, in the energy band,  
from  $3 \times 10^{18}$ to $5 \times 10^{19}$ $eV$ an intensity maximum  is expected,
being the ankle energy for protons at about 
$ 3 \times 10^{18}$ $eV$, a bit lower than that of helium at 
$ 7 \times 10^{18}$ $eV$. Accurate measurements of the Hires Collaboration 
in the energy band $10^{18}$-$10^{20}$ $eV$ 
do actually indicate an intensity bump in this region (Springer et al., 2003). Other
experiments (Haverah park, Yakutsk, Agasa) register a similar intensity bump
with respect to the extrapolation of the spectrum, at constant spectral index,
measured at lower energies, below $ 10^{18}$ $eV$. This
inference and the conclusion for the existence of an intensity bump are independent, 
within plausible empirical limits, from the ion blend and the spectral indices of 
the extragalactic component, and to a large extent, from the method of
calculation.

%%%%% ANCIENNE PLACE DES TROIS FIGURES 15 16 17

\section{Conclusions}

The uneven efficiency
of  the magnetic field to retain cosmic rays in the Galaxy at high energies,
 above  $10^{13}$ $eV$,    is insufficient
to account for the knees of the individual ions and for the knee. This
decreasing efficiency would yield 
an enhancement or a depression  
of the flux at Earth, 
depending on the nuclear species and the particular energy. 
It is only by the detailed analysis of the effects of nuclear cross sections,
the position of the solar cavity and the disc size on cosmic ray flux
at Earth that the origin of the knees and ankles becomes fully comprehensible. 
Figure 17 shows the number of galactic iron 
nuclei $n_g$ crossing the disc border (top curve) and that
 of galactic iron nuclei reaching the Earth (bottom curve) versus
energy, exactly in the same conditions. The break of $n_g$ 
caused by the magnetic
field occurs, in both cases,  precisely around and above $ 10^{17}$ $eV$, but 
the effect on the iron intensity
measurable by  a terrestrial or a peripheral instrument
is opposite: in one case there is a knee, in the other an antiknee. 
This behavior vividly illustrates
the importance of the position of the Earth within the Galaxy.
 
\par Let us summarize the features of the solution of the knee and ankle
 problem presented at this Conference : 
 
(1)  the energy spectra of proton,  helium and iron nuclei fairly well 
     match those measured 
     by the Kascade and Eas-top experiments; 
(2)  the computed knee of the complete spectrum is also in agreement
     with the experimental data; this agreement is substantiated
     by the correct bend energy, between  $3-5 \times 10^{15}$ $eV$,   
     and the correct spectral index of 3 
     above $6 \times 10^{15}$ $eV$  (see figures 11 and 12 and their legends). 
(3)  The positions of the ankles and knees along the energy axis are distinctively
     and uniquely interconnected by the average magnetic field strength,  which forges
     the grammage, and by the rate at which inelastic cross sections rise with energy,
     as shown in figure 15. This fundamental conclusion corroborates the 
     explanation of the knees and  ankles described in Section 3 and 4.

\par It seems to us that the coherent set of  observational 
     facts and related physical processes, affecting the propagation
     of galactic cosmic rays in the disk, as described in Sections 2 and 3,
     form the irreducible basis for any further 
     quantitative exploration
     of the knees and ankles of the cosmic radiation.
This quantitative exploration
is probably feasible using other simulation codes (see, for example,
Moskalenko, 1997).

\end{document}